# Symmetry Energy and the Isoscaling in Reactions on Enriched Tin Isotopes

A.S.Danagulyan, A.R.Balabekyan, G.H.Hovhannisyan

*Yerevan State University, Alex Manoogian 1, 0025, Armenia*

Corresponding author: A.S.Danagulyan   danag@ysu.am

The coefficients of symmetry energy term for fragments with Z=4,11,12 measured in multifragmentation reactions initiated by proton and deuteron with energy of 3.65A GeV on enriched tin isotopes $^{112,118,120,124}$Sn are determined. The dependence of isoscaling parameter on the excitation energy, the temperature of fragmenting systems and the density ratio for heavy mass products are analyzed.

## 1. Introduction

In the last few years of large interest have been investigations of isoscaling phenomenon in nuclear reactions in liquid-gas phase transition region. This is accounted for by the use of isoscaling parameter in obtaining information about symmetry energy in nuclear equation-of-state (EOS) in regions away from normal density conditions [1-7]. Besides the nuclear point of view such investigations are also of great importance for understanding some processes in astrophysics, in particular, in supernovae explosions and neutron stars [8-11].

The isotopic effect was observed by Bogatin *et al.* for reactions induced by protons, deuterons and $\alpha$ particles of incident energies between 660 MeV and 15.3 GeV on $^{112,124}$Sn targets [12-14]. The dependence of yields of photonuclear reactions and nucleon-nuclear reactions on values of third projection of isospin of target and product nuclei was investigated in our previous works [15-18].

In a series of recent papers the scaling properties of cross sections for fragment production with respect to the isotopic composition of the emitting systems were investigated by Tsang et al. [19-21]. It has been shown, that the yield ratio of a given isotope produced in two reactions with different isospin asymmetry exhibits an exponential dependence on proton and neutron number. This observation known as isoscaling has been identified in a variety of reaction mechanisms, including multifragmentation process:

$$R_{21} = Y_2(N,Z)/Y_1(N,Z) = C\, exp(\alpha N + \beta Z) , \qquad (1)$$

where $Y(N,Z)$ is the yield of a fragment with $Z$ protons and $N$ neutrons. Indices "1" and "2" correspond to different targets with different isotopic compositions, with "2" corresponding to more neutron-rich target. $C$ is a normalization coefficient. The parameters $\alpha$ and $\beta$ were expressed using the difference of chemical potentials of the two systems as follows: $\alpha = \Delta\mu_n/T$, and $\beta = \Delta\mu_p/T$ [20], where $T$ is the temperature of the excited nucleus.

Based on the statistical interpretation of isoscaling, the coefficient of the symmetry–term in the nuclear mass can be extracted using isoscaling parameter [7, 22].

The aim of this paper is to obtain the coefficient of symmetry energy term in multifragmentation reactions by using the values of isoscaling parameters for fragments in Z=4,11,12 charge region in proton and deuteron induced reactions with energy of beams 3.65A GeV on enriched tin targets ($^{112,118,120,124}$Sn). The temperature of the fragmenting system has been determined by untraditional methods.

## 2. Experimental details and discussion.

The targets of tin isotopes $^{112,118,120,124}$Sn (enrichments - 92.6%, 98.7%, 99.6%, 95.9%, respectively) were irradiated at the Nuclotron and Synchrotron of the JINR (Joint Institute for Nuclear Research at Dubna) by proton and deuteron beams with the energy of 3.65 GeV/nucleon. The description of the experiment is given in [23].

**Isoscaling.** In our previous studies of the nuclear reactions induced by protons with energies of 0.66 and 8.1, 3.65 GeV, deuterons (E=3.65 AGeV) and $^{12}$C ions (E=2.2 A GeV) on targets of tin isotopes the isoscaling behavior was shown using the induced activity method [17,18] and it was described in terms of the third component of the fragment isospin $t_3$=(N-Z)/2. The isotopic ratio of the product formation cross sections has been considered in the following form:

$$R_{21}=Y_2(N,Z)/Y_1(N,Z)=exp(C+Bt_3), \qquad (2)$$

where *C* and *B* are fitting parameters. The parameter *B* is related to the difference of the chemical potentials of neutrons of the two fragmenting systems. ($B=2\alpha$, where $\alpha$ is a coefficient in formula (1)). A similar form of description of isotopic dependency of the yields of products in the reactions initiated by protons and light ions with energies of *0.66 GeV* to *15.3 GeV* on the separated tin isotopes was used in [7].

In this work we discuss multifragmentation products with *A=7-28* to find the symmetry coefficient *γ*. The discussions of experimental values of cross section ratios of fragments produced in reactions induced by protons and deuterons with the *3.65GeV/nucleon* incident energy on enriched tin isotopes affirmed the existence of isoscaling.

Fig.1 shows the isoscaling exhibited by fragments $^7$Be, $^{22}$Na, $^{24}$Na, $^{28}$Mg measured in the reactions of *p* and *d* + $^{112,118,120,124}$Sn.

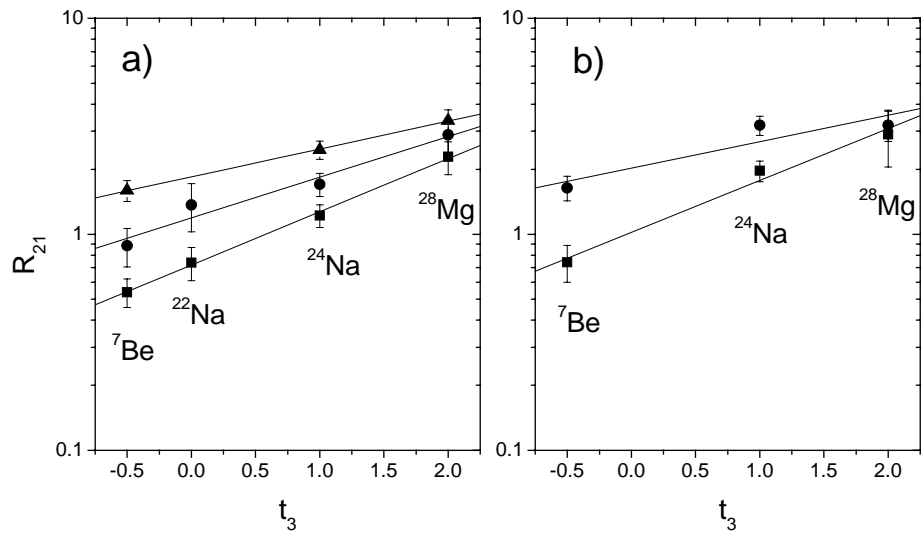

Figure 1. Yield ratio $R_{21}$ versus the third component of the fragment isospin $t_3$ for different target pairs bombarded by protons (a) and deuterons (b). Symbols (■) for target pairs $^{124}$Sn /$^{112}$Sn; (●) for target pairs $^{120}$Sn /$^{112}$Sn ($R_{21}\times 1.5$ for p and $R_{21}\times 2$ for d); (▲) for target pairs $^{124}$Sn /$^{118}$Sn ($R_{21}\times 2$ for p). The lines are the result of exponential fits according to Eq.(2).

The values of fitting parameter $B$ obtained from experimental data (for target pairs $^{124}Sn/^{112}Sn$) are $0.57\pm0.02$ and $0.55\pm0.07$ for proton and deuteron induced reactions respectively, which is in agreement with the values extracted from other multifragmentation reactions [19-21].

## 3. Excitation energy, temperature, symmetry energy

In the recent years the interest in isoscaling of the reaction products has considerably increased, motivated by the possibility to extract information on the symmetry energy of hot nuclei and nuclear matter during the liquid –gas phase transition.

Investigation of isoscaling allows us to obtain the symmetry energy coefficient $\gamma$ in the symmetry term in the equation of state. It has been shown [7,20], that the symmetry energy in the statistical model calculations is related to the isoscaling parameter $\alpha$ ($B=2\alpha$) through the relation

$$\alpha T = 4\gamma(Z_1^2/A_1^2 - Z_2^2/A_2^2) , \qquad (3)$$

where $\alpha$- is the isoscaling parameter for the hot primary fragments, i.e. before their sequential decay into cold secondary fragments; $Z_1, A_1$ and $Z_2, A_2$ are the charges and mass numbers of the two fragmenting systems; $T$- is the temperature of the systems; $\gamma$- is the symmetry energy coefficient.

The temperature $T$ is generally obtained from the double ratios of cross sections of two light isotopes formed in multifragmentation process, such as $^{3,4}He / ^{6,7}Li$ , $^{4,6}He / ^{6,8}Li$ , $^{6,8}He / ^{6,8}Li$ [7].

In the present work we use the expanding Fermi gas model predictions according to which dependence of temperature $T$ on the excitation energy $E^*$ of emitting source is the following [1 and references therein]:

$$T = [K_0 (\rho/\rho_0)^{0.69} E^*]^{0.5} , \qquad (4)$$

where $K_0 = 10$ is the inversed level density parameter. The value of degree for density ratio is taken from [1,2].

The excitation energy of the source for each beam was determined using the Standard Two-Step vector model. We determine the recoil properties of nuclei with "thick-target thick-catcher" experiment using the induced activity method [24]. The quantities measured were the fractions F and B of product nuclei that recoil out of the target ($^{118}Sn$) foil in forward and backward directions related to the beam. The value of the forward velocity $v_{\parallel}$ was determined, which can be used to obtain the average cascade deposition energy (excitation energy $E^*$). The relation between the excitation energy $E^*$ and value of the forward velocity $v_{\parallel}$ can be estimated as

$$E^* = 3.253 \cdot 10^{-2} \cdot k' A_t v_{\parallel} (T_p/(T_p+2))^{0.5} , \qquad (5)$$

where excitation energy $E^*$ and bombarding energy $T_p$ are expressed in terms of $m_p c^2$, $A_t$ is the target mass in *amu* and $v_{\parallel}$ is in units of $(MeV/amu)^{1/2}$. The constant $k'$ has been evaluated by O.Scheidemann and N.Poril [25] on the basis of Monte-Carlo cascade calculations as $k'=0.8$.

A mean excitation energy for light isotopes $^{24}Na$ and $^{28}Mg$ is *591,5MeV* and *535.4 MeV* for *p* and *d* induced reactions respectively[24]. According to INC calculations the residual mass number of the system after cascade process $A_{res} \approx 106$ for the $^{118}Sn$ target, since multiplicity of the emitted particles at energy $Ep=3.65GeV$ is equal 13 ($\Delta A=n_p+n_n=13$), and $A_{res}=A_t+1(2)-\Delta A$) [26]. Mean excitation energy per nucleon is $\bar{E}^*/A_{res} \approx 5.58\ MeV$.

For deduced density ratio $\rho/\rho_0$ we used excitation energy $E^*$ dependence of the density $\rho/\rho_0$ obtained for reactions with fragmenting system mass $A\approx100$ in [1]. In that work on Fig.10 experimental points obtained for $^{58}Fe+^{58}Fe$, $^{58}Fe+^{58}Ni$ and $^{58}Ni+^{58}Ni$ reactions at beam energies of *30, 40* and *47 A MeV* are shown. For comparison, there are also shown the break-up densities obtained from the analysis of the apparent level density parameters required to fit the measure caloric curve by J.Natowitz et al. [27], those obtained by V.Viola et al. [28] from the Coulomb barrier systematic that are required to fit the measured intermediate mass fragment kinetic energy spectra and calculation dates of the SMM model [29]. All discussed results are in good agreement.

For $\bar{E}^*/A_{res} \approx 5.58$ MeV the density ratio is determined to be $\rho/\rho_0 = 0.5$ which indicates liquid-gas transition process. This value of density ratio agrees with that in Antisymmetrized Molecular Dynamics (AMD). According to (AMD) model the liquid-gas transition begins when $\rho/\rho_0 = 0.08/0.16 = 0.5$. The fragments are formed when $\rho \sim 0.08$ fm$^{-3}$ [22].

Furthermore, having values of excitation energy $E^*$ and density ratio $\rho/\rho_0$ for proton and deuteron induced reactions, we can calculate the temperature $T$ from formula (5). We obtained $T = 5.9$ MeV and $5.59$ MeV for proton and deuteron induced reactions respectively.

Using the experimental values of isoscaling parameter $\alpha$ and values of temperature $T$ calculated with the above mentioned method, we estimated the symmetry coefficient $\gamma_{app}$ from formula (3) for systems with various isotopic compositions. We discuss the following target pairs: ($^{124}$Sn/$^{112}$Sn), ($^{120}$Sn/$^{112}$Sn), ($^{124}$Sn/$^{118}$Sn).

The obtained data of excitation energy per nucleon $\bar{E}^*/A_{res}$ of fragmenting system, isoscaling parameter $\alpha$, density ratio $\rho/\rho_0$ and temperature $T$ are brought in Table 1 for proton and deuteron induced reactions.

Table 1
Average values of $\bar{E}^*/A_{res}$, $\rho/\rho_0$, $\alpha = B/2$, $T$ and $\gamma_{app}$ for product nuclei produced in with 3.65 GeV/nucleon incident energy.

| Type of reactions | $\bar{E}^*/A_{res}$ | $\rho/\rho_0$ | $T$ (MeV) | ($\Delta N=12$) $\alpha=B/2$ | ($\Delta N=8$) $\alpha=B/2$ | ($\Delta N=6$) $\alpha=B/2$ | ($\Delta N=12$) $\gamma_{app}$(MeV) | ($\Delta N=8$) $\gamma_{app}$(MeV) | ($\Delta N=6$) $\gamma_{app}$(MeV) |
|---|---|---|---|---|---|---|---|---|---|
| proton induced reaction | 5.58 | 0.50 | 5.90 | 0.285 | 0.215 | 0.145 | 11.45 | 12.3 | 12.6 |
| deuteron induced reaction | 5.05 | 0.50 | 5.59 | 0.275 | – | 0.13 | 10.48 | – | 10.71 |

For various isotopic compositions of the two systems isoscaling coefficients are different, but values of $\gamma_{app}$ are nearly the same.

We estimate $\gamma_{app} \approx 10$-$13$ MeV by applying formula (3) to the experimentally determined $\alpha$ (Tab.1). Errors of $\gamma_{app}$ are about 15-20%, provided by the errors of the forward velocity $v_{\parallel}$ and $E^*$ calculations with Two-Step Vector model [24], by $\rho/\rho_0$ calculations from SMM model and the experimental data [1]. Obtained value of $\gamma_{app}$ is significantly lower than the value of $\gamma = 25$ MeV for isolated cold nuclei in ground state at normal density.

It should be mentioned that in the statistical approach, the formula (3) was obtained for the freeze-out conditions. To establish the connection between coefficients of symmetry of hot fragments and those obtained for the observed cold fragments we should take into account the process of secondary deexcitation. The values of the symmetry coefficient for light nuclei will slightly increase for primary processes with about 5-8% according to [20] or 20% according to [7].

However, as the SMM calculations for fragments with $Z \geq 10$, for which secondary-deexcitation is described essentially as a sequential evaporation, show the experimental $\gamma_{app}$ may be even larger than the $\gamma$ of the fragments in the freeze-out [3-6]. In the article by D.Henzelova et al. [6] results of measurement of $^{136}$Xe and $^{124}$Xe projectiles in peripheral to mid-peripheral collisions with a lead target at 1 A GeV were discussed within the statistical approach describing the liquid-gas nuclear phase transition. The residues produced in these reactions were identified with the use of the Fragment Separator at GSI, Darmstadt. Inclusive fragment yields are measured. In this studies the authors analyze the isoscaling phenomenon and symmetry energy of fragments with Z=10-13. It is important, that the region of considered fragments is similar with that investigated by us (Z=11,12) and our targets asymmetry ($^{124}$Sn, $^{112}$Sn) coincide with the projectiles asymmetry in [6]. Thus, comparison with our data is reasonable.

Carrying out SMM calculations taking into account secondary deexcitation of fragments and using both the <N/Z> ratio and the isoscaling for the purpose of investigating the symmetry energy coefficient authors concluded that $\gamma \approx 11$-$15$ MeV rather well reproduced the experimental data [6]. The observed decrease of symmetry coefficient is consistent with the previous investigation of multifragmentation [3-5]. The value of symmetry coefficient $\gamma_{app}$ obtained from our data is in good agreement with this conclusion.

It is possible to calculate values of $\rho/\rho_0$ and $T$ for heavier product nuclei produced in proton and deuteron induced reaction with *3.65 GeV/nucleon* incident energy using above mentioned procedure. The values of $\bar{E}^*$ have been taken from [24] and isoscaling parameter $\alpha=B/2$ from [18]. Obtained results are given in Tab.2,3.

Table 2

Average values of $\bar{E}^*/A_{res}$, $\alpha$, $\rho/\rho_0$, $T$ for product nuclei produced in proton induced reaction with *3.65 GeV* incident energy.

| Mass region | $\bar{E}^*/A_{res}$ | $\alpha=B/2$ | $\rho/\rho_0$ | T (MeV) |
|---|---|---|---|---|
| 42≤A≤58 | 3.86 | 0.37 | 0.70 | 5.49 |
| 67≤A≤77 | 2.89 | 0.42 | 0.80 | 4.97 |
| 81≤A≤89 | 2.72 | 0.43 | 0.85 | 4.93 |
| 90≤A≤99 | 1.70 | 0.44 | 0.98 | 4.10 |
| 104≤A≤109 | 0.77 | 0.67 | 1 | 2.77 |

Table 3

Average values of $\bar{E}^*/A_{res}$, $\alpha$, $\rho/\rho_0$, $T$ for product nuclei produced in deuteron induced reaction with *3.65 GeV/nucleon* incident energy.

| Mass region | $\bar{E}^*/A_{res}$ | $\alpha=B/2$ | $\rho/\rho_0$ | T (MeV) |
|---|---|---|---|---|
| 42≤A≤58 | 3.34 | 0.37 | 0.74 | 5.21 |
| 67≤A≤77 | 2.45 | 0.40 | 0.89 | 4.75 |
| 81≤A≤89 | 2.39 | 0.435 | 0.90 | 4.72 |
| 90≤A≤99 | 1.28 | 0.45 | 0.99 | 3.56 |
| 104≤A≤109 | 0.86 | 0.49 | 1 | 2.92 |

As can be seen from Tables 2 and 3, isoscaling parameter $\alpha=$ B/2 and ratio $\rho/\rho_0$ increases, whereas $\bar{E}^*/A_{res}$ and T decreases with increase of the product mass number independently of projectile sort. Apparently this is connected with different mechanisms of production of these nuclei. The heavy products (with mass near of target) are more likely to be produced in spallation reactions. The low mean values of parameter B for light nuclei (A=7,24,28) (see Tab.1) are characteristic for multifragmentation mechanisms. In formation process of the products in mass region between these two ranges, possibly, there are contributions from various channels of production, i.e. both spallation and multifragmentation. This conclusion has been done by analyses of the values of isoscaling parameters. [18].

**Conclusions**

To sum up, we have determined the coefficients of symmetry energy term of the fragments with Z=4,11,12 measured in multifragmentation reactions by proton and deuteron with energy 3.65 AGeV on enriched tin isotopes $^{112,118,120,124}$Sn using isoscaling parameters of systems with different asymmetry. We have obtained the values $\gamma_{app}=11,5 MeV$ and $10,5 MeV$ for proton and deuteron induced reactions, respectively. For temperature determination we use the expanding Fermi gas model predictions. We conclude that $\gamma_{app}$ does not depend on target pairs' asymmetry, within the limits of errors. For heavy mass products it is observed that the isoscaling parameter $\alpha$ or $B/2$ decreases with increasing excitation energy and temperature of fragmenting systems, and increases with the increase of density ratio $\rho/\rho_0$.

**Acknowledgment.** We would like to express our gratitude to the second year Master student A.G.Movsesyan for help in preparing this paper.

**REFERENCES**

[1] D.V.Shetty, S.J.Yennello and G.A. Souliotis**,** Phys.Rev.C 76, (2007) 024606,
[2] G.A.Souliotis,A.S.Botvina,D.V.Shetty,et al. Phys.Rev.C 75,(2007) 011601(R)
[3]  J.Iglio,D.V.Shetty,S.J.Yennello, et al. Phys. Rev. C **74**,024605 (2006)
[4] D.V.Shetty, A.S.Botvina,S.J.Yennello, et al. Phys. Rev. C **71**,024602 (2005)
[5] A.Le Fevre, G.Auger, M.L.Begemann-Blaich, *et al.*, Phys. Rev. Letters, PRL **94**, (2005), 162701.
[6] D.Henzelova,A.S.Botvina,K.H.Schmidt, et al. ArXiv nucl-ex/0507003(2005)
[7] A.S.Botvina, O.V.Lozhkin, W.Trautmann, Phys. Rev. C, 65, (2002), 044610.
[8] A.S.Botvina and I.N. Mishustin, Phys.Lett. B585,233(2004)
[9] J.M.Lattimer and M.Prakash,Astrophys. J.550,426(2001)
[10] C.J.Pethick and D.G.Ravenhall,Annu. Rev.Nucl.Part.Sci. 45,429 (1995)
[11] A.S.Botvina and I.N. Mishustin, Phys.Rev.C 72,048801(2005)
[12]V.I.Bogatin,V.K.Bondarev,V.F.Litvin, et al.Yad.Fiz.19,32,(1974),Sov.J.Nucl.Phys.19,16,(1974)
[13] V.I.Bogatin,E.A.Ganza,O.V.Lozhkin,et al. Yad.Fiz.31,845(1980),Sov.J.Nucl.Phys.,31,436(1980), Yad.Fiz.34,104(1981),Sov.J.Nucl.Phys.34,59(1981)
[14] V.I.Bogatin,E.A.Ganza,O.V.Lozhkin,et al. Yad.Fiz.36,33(1982),Sov.J.Nucl.Phys.,36,33(1982)
[15] G.V.Arustamyan, H.H.Vartapetyan, A.S.Danagulyan, *et al.*,  Yad. Fiz., **32**, (1980), 1165;
A.A.Arakelyan,G.A.Arustamyan,A.S.Danagulyan, *et al.*, Yad. Fiz., **35**, (1982), 518
[16] V.E.Aleksadrian,A.A.Arakelyan,A.S.Danagulyan, Yad.Fiz., **57**, (1994), 2115
[17] A.R.Balabekyan, A.S.Danagulyan, J.R.Drnoyan *et al.*, Nucl.Phys.A **735**, (2004), 267
[18] A.S.Danagulyan, A.R.Balabekyan,G.H.Hovhannisyan,et al.Nucl.Phys.A 814,(2008), 109
[19] M.B.Tsang, W.A.Friedman, C.K.Gelbke , *et al.*, Phys. Rev. C **64**, (2001), 041603.( R )
[20] M.B.Tsang, C.K.Gelbke, X.D.Liu *et al.*,  Phys.Rev.C **64**, (2001), 054615.
[21] M.B.Tsang, W.A.Friedman, C.K.Gelbke , *et al.*,  Phys.Rev.Lett. **86**, (2001), 5023.
[22] A.Ono,P.Danielewicz,W.Friedman, *et al.*, Phys. Rev. C **68**, (2003), 051601
[23] A.R.Balabekyan, A.S.Danagulyan, J.R.Drnoyan *et al.*, Yad. Fiz., **69**, (2006), 1520, Physics of Atomic Nuclei, **69**, (2006), 1485
 [24]. A..Balabekyan, A..Danagulyan, J.Drnoyan *et al.* Yad. Fiz.,**70,** (2007),1940, Physics of Atomic Nuclei, **70**, (2007), 1889.
[25] O.Scheidemann and N.Porile, Phys.Rev.C 14,1534 (1976)
[26]  V.S.Barashenkov, V.D.Toneev, Interactions of High Energy Particles and Atomic Nuclei  with Nuclei. Moscow, 1972, p.358 (in Russian).
[27]J.B.Natowitz,R.Wada,K.Hagel, at al., Phys.Rev.C 66, 031601 (2002)
[28] V.E.Viola ,K.Kwiatkowski,J.B.Natowits,et al., Phys.Rev. Lett. 93, 132701 (2004)
[29] J.P.Bondorf,R.Donangelo,I.N.Mishustin and H.Schultz, Nucl.Phys.A 444,460 (1985)
[30] J.P.Bondorf,A.S.Botvina,A.S.Iljinov,et al.Phys.Rep. 257,133 (1995)
[31] A.S.Botvina and I.N.Mishustin,Phys.Rev.C 63,0610601 (2001)